\documentclass[aip,jcp,amsmath,amssymb,citeautoscript,preprint]{revtex4-1}

\usepackage{graphicx}
\usepackage{dcolumn}
\usepackage{bm}

\usepackage[utf8]{inputenc}
\usepackage[T1]{fontenc}
\usepackage{mathptmx}

\usepackage{longtable}
\usepackage{footnote}
\usepackage{float}

\usepackage{color}

\begin{document}

\newcommand{\p}{^\prime}
\newcommand{\pp}{^{\prime\prime}}
\newcommand{\ai}{{\it ab initio}}
\newcommand{\cm}{cm$^{-1}$}
\def\a0{{$a_{\rm 0}$}}
\newcommand{\X}{$\tilde{X}\,^2\Sigma^+$}
\newcommand{\A}{$\tilde{A}\,^2\Pi$}

\title{Theoretical rovibronic spectroscopy of the calcium monohydroxide radical (CaOH)}

\author{Alec Owens}
\email{alec.owens.13@ucl.ac.uk}
\affiliation{Department of Physics and Astronomy, University College London, Gower Street, WC1E 6BT London, United Kingdom}

\author{Victoria H. J. Clark}
\email{v.clark.17@ucl.ac.uk}
\affiliation{Department of Physics and Astronomy, University College London, Gower Street, WC1E 6BT London, United Kingdom}

\author{Alexander Mitrushchenkov}
\email{Alexander.Mitrushchenkov@univ-eiffel.fr}
\affiliation{MSME, Univ Gustave Eiffel, CNRS UMR 8208, Univ Paris Est Creteil, F-77474 Marne-la- Vallée, France}

\author{Sergei N. Yurchenko}
\email{s.yurchenko@ucl.ac.uk}
\affiliation{Department of Physics and Astronomy, University College London, Gower Street, WC1E 6BT London, United Kingdom}

\author{Jonathan Tennyson}
\email{j.tennyson@ucl.ac.uk}
\affiliation{Department of Physics and Astronomy, University College London, Gower Street, WC1E 6BT London, United Kingdom}

\date{\today}

\begin{abstract}
The rovibronic (rotation-vibration-electronic) spectrum of the calcium monohydroxide radical (CaOH) is of interest to studies of exoplanet atmospheres and ultracold molecules. Here, we theoretically investigate the \A--\X\ band system of CaOH using high-level \textit{ab initio} theory and variational nuclear motion calculations. New potential energy surfaces (PESs) are constructed for the \X\ and \A\ electronic states along with $\tilde{A}$--$\tilde{X}$ transition dipole moment surfaces (DMSs). For the ground \X\ state, a published high-level \textit{ab initio} PES is empirically refined to all available experimental rovibrational energy levels up to $J=15.5$, reproducing the observed term values with a root-mean-square (rms) error of 0.06~cm$^{-1}$. Large-scale multireference configuration interaction (MRCI) calculations using quintuple-zeta quality basis sets are employed to generate the \A\ state PESs and $\tilde{A}$--$\tilde{X}$ DMSs. Variational calculations consider both Renner-Teller and spin-orbit coupling effects, which are essential for a correct description of the spectrum of CaOH. Computed rovibronic energy levels of the \A\ state, line list calculations up to $J=125.5$, and an analysis of Renner-Teller splittings in the $\nu_2$ bending mode of CaOH are discussed.
\end{abstract}

\maketitle

\section{Introduction}
\label{sec:intro}

Recently, a number of research fields have shown increasing interest in the rovibronic spectrum of the calcium monohydroxide radical ($^{40}$Ca$^{16}$O$^{1}$H). CaOH is expected to occur in the atmospheres of hot rocky super-Earth exoplanets but a lack of molecular line list data is hampering its detection~\citep{09Bernath.exo,jt693}. This type of exoplanet is in close proximity to its host star and subject to extremely high temperatures, e.g.\ 2000--4000~K, which causes the material on the surface of the planet to vaporise. The result is an atmosphere strongly dependent on planetary composition where a number of simple molecules composed of rock-forming elements (Si, Ca, Mg, Fe) are anticipated~\citep{12ScLoFe.exo,16FeJaWi}. A recent systematic study modelling M-dwarf photospheres also noted missing opacity from the benchmark BT-Settl model due to CaOH, notably its band around 18\,000~cm$^{-1}$~\citep{13RaReAl.CaOH}, and CaOH has been included in models of hot-Jupiter exoplanet atmospheres~\citep{15ArWaxx.CaOH}. The fields of ultracold molecules and precision tests of fundamental physics have been focusing on the alkaline earth monohydroxide radicals, notably CaOH~\citep{19KoStYu.CaOH,19AuBoxx.CaOH}. The rovibronic energy level structure is particularly favourable when designing efficient laser cooling schemes and direct laser cooling of CaOH to temperatures near 1~mK has already been demonstrated in a one-dimensional magneto-optical trap~\citep{20BaViHa.CaOH}. Experimental schemes to extend cooling of CaOH into the microkelvin regime are actively being considered~\citep{21BaViHa.CaOH}.

The calcium monohydroxide radical is an open-shell, linear molecule with a relatively complex electronic structure. Measurements of its rovibronic (rotation-vibration-electronic) spectrum~\citep{06DiShWa.CaOH,97HaJaBe.CaOH,96PeLexx.CaOH,96ZiFlAn.CaOH,96LiCoxx.CaOH,95LiCoxx.CaOH,94CoLiPr.CaOH,93ScFlSt.CaOH,92LiCoxx.CaOH,92CoLiPr.CaOH,92ZiBaAn.CaOH,92JaBexx.CaOH,91CoLiPr.CaOH,85BeBrxx.CaOH,84BeKixx.CaOH,83HiQiHa.CaOH} have revealed the lowest-lying eight electronic states up to the $\tilde{G}\,^2\Pi$ state at approximately 32\,633~cm$^{-1}$~\citep{97HaJaBe.CaOH}. Interestingly, the spectrum of CaOH is modified by the Renner-Teller effect~\citep{Renner1934}, which occurs in molecules with a doubly degenerate electronic state at linearity, e.g. the $\tilde{A}\,^2\Pi$ state of CaOH. As the molecule bends, the interaction of the electronic orbital angular momentum and the nuclear vibrational angular momentum lifts the degeneracy of the electronic state, which splits into two non-degenerate components. The Renner-Teller effect significantly complicates spectral analysis and must be considered in any correct description of CaOH spectra~\citep{96LiCoxx.CaOH,95LiCoxx.CaOH,94CoLiPr.CaOH,92LiCoxx.CaOH,92JaBexx.CaOH,91CoLiPr.CaOH,83HiQiHa.CaOH}.

There have been a number of theoretical studies on the electronic structure of CaOH~\citep{84BaPaxx.CaOH,86BaLaPa.CaOH,90BaLaSt.CaOH,90Ortiz.CaOH,96KoBoxx.CaOH,02KoPexx.CaOH,02ThPeLi.CaOH,05TaChFr.CaOH}, but the majority of these have focused on molecular structures and properties. The most relevant work to that presented here is of Koput and Peterson~\citep{02KoPexx.CaOH}, who computed an accurate \textit{ab initio} potential energy surface for the ground \X\ state using the spin-restricted coupled cluster method, RCCSD(T), with large correlation consistent basis sets of quintuple-zeta quality. Variational rovibrational energy level calculations showed excellent agreement with experiment, reproducing the known fundamental wavenumbers (the Ca--O stretching $\nu_1$ and bending $\nu_2$ mode) to within $1$~cm$^{-1}$ and excited vibrational levels to within $4$~cm$^{-1}$. 

In this work, we computationally investigate the \A--\X\ rovibronic spectrum of CaOH, considering both Renner-Teller and spin-orbit coupling effects in our spectroscopic model. New potential energy surfaces (PESs) for the \X\ and \A\ states (which splits into two Renner surfaces $A^{\prime}$ and $A^{\prime\prime}$ at bent configurations of CaOH) are determined along with \A--\X\ transition dipole moment surfaces (DMSs). Variational nuclear motion calculations are performed with the EVEREST code~\citep{EVEREST}, which employs an exact kinetic energy operator and is capable of treating an arbitrary number of electronic states with strong Renner-Teller coupling. Rovibrational and rovibronic energy levels of CaOH are calculated and compared against available experimental data to validate our spectroscopic model. A preliminary molecular line list covering rotational excitation up to $J=125.5$ in the 0--30\,000~cm$^{-1}$ range is generated and Renner-Teller splittings in the $\nu_2$ bending mode of CaOH are discussed.

\section{Computational details}
\label{sec:methods}

\subsection{Potential Energy Surfaces}


The ground and excited state PESs were each represented using the analytic function,
\begin{equation}
\label{eq:pot}
V =  \sum_{ijk} f_{ijk} \xi_1^{i} \xi_2^{j} \xi_3^{k},
\end{equation}
with maximum expansion order $i+j+k=8$. The vibrational coordinates,
\begin{eqnarray}
\label{eq:coords_pes}
  \xi_1 &=& (r_1-r_1^{\rm eq})/r_1, \\
  \xi_2 &=& (r_2-r_2^{\rm eq})/r_2, \\
  \xi_3 &=& \alpha-\alpha_{\rm eq},
\end{eqnarray}
where the internal stretching coordinates $r_1  = r_{\rm CaO} $, $r_2  = r_{\rm OH} $, the interbond angle $\alpha = \angle({\rm CaOH})$, and the equilibrium parameters are $r_1^{\rm eq}$, $r_2^{\rm eq}$ and $\alpha_{\rm eq}$ (see supplementary material for values). Note that the exponent $k$ associated with the bending coordinate $\xi_3$ assumes only even values because of the symmetry of CaOH. 

For the ground \X\ state, we have taken the original \ai\ PES of Koput and Peterson~\citep{02KoPexx.CaOH} and empirically refined it to term values up to $J=15.5$ using an efficient least-squares fitting procedure~\citep{jt503} in the variational nuclear motion program TROVE~\citep{jt466}. The data used in the refinement for the \X\ state was taken from a recent MARVEL (measured active rotation vibration energy levels) analysis of the published spectroscopic literature of CaOH~\citep{jt791}, where a robust algorithm was able to provide 1955 experimental-quality levels for the five lowest-lying electronic states. The relevant data for this study covered the regions 0--2599~cm$^{-1}$ ($\tilde{X}\,^2\Sigma^+$) and 15\,966--17\,677~cm$^{-1}$ ($\tilde{A}\,^2\Pi$). In the refinement, only five of the ground state expansion parameters $f_{ijk}$ were varied (the three quadratic parameters plus the quartic and sextic bending parameters), along with the equilibrium parameters $r_1^{\rm eq}$ and $r_2^{\rm eq}$. Since TROVE is unable to treat spin interactions (as opposed to the program EVEREST), spin-unresolved experimental term values were used in the refinement. These were formed by taking the average of the two separate spin component energies for each rovibrational state. Pure rotational energies were given a higher weighting in the refinement as they had smaller measurement uncertainties compared to excited vibrational states. Atomic mass values of 39.951619264818~Da (Ca), 15.990525980297~Da (O), and 1.007276452321~Da (H)~\citep{AME_2012} were used in the TROVE refinement, and these values were subsequently used in rovibronic calculations with EVEREST (discussed in Section~\ref{sec:everest}) to ensure the accuracy of the refined \X\ PES was maintained. The results of the \X\ state refinement will be discussed in Section~\ref{sec:results}.

For the first excited \A\ state, electronic structure calculations were carried out on a grid of approximately 2700 nuclear geometries in the range $h c \cdot 15\,377$--35\,000~cm$^{-1}$ ($h$ is the Planck constant and $c$ is the speed of light) using the quantum chemical program MOLPRO2015~\citep{MOLPRO,Molpro:JCP:2020}. Initially, state-averaged multi-configurational self-consistent field (MCSCF) calculations~\citep{85KnWexx.ai,85WeKnxx.ai} including the ground \X\ and excited \A\ (both $A^{\prime}$ and $A^{\prime\prime}$ components) states were performed. The resulting molecular orbitals were used in internally-contracted multireference configuration interaction (MRCI) calculations in conjunction with the correlation consistent basis sets cc-pV5Z for Ca~\citep{02KoPexx.CaOH}, and aug-cc-pV5Z for O and H~\citep{92KeDuHa.ai}. The active space included 7 electrons distributed between $(10\,a^{\p},5\,a^{\pp})$ orbitals in $\bm{C}_{\mathrm{s}}$ point group symmetry. This choice of active space yielded smooth surfaces for the $A^{\prime}$ and $A^{\prime\prime}$ components and the correct Renner-Teller behaviour, i.e.\ degenerate electronic energies for the two components at linear geometries. Previous computational studies of the low-lying electronic states of CaOH~\citep{02ThPeLi.CaOH,05TaChFr.CaOH} have investigated different choices of active space but these studies were concerned with many more electronic states than are required here.

Discussing the quality of our electronic structure calculations, there are of course errors arising from our treatment. The largest available correlation consistent basis sets for Ca are of quintuple-zeta quality and we have utilised the original cc-pV5Z basis set~\citep{02KoPexx.CaOH}, which was used to produce an accurate \textit{ab initio} ground state PES of CaOH. Note that more recent pseudopotential-based correlation consistent basis sets up to quintuple-zeta quality have been developed for Ca~\citep{17HiPexx.ai}. Schemes exist to remove the error associated with truncation of the one-particle basis set through extrapolation to the complete basis set limit, but we believe this incompleteness error to be small enough to neglect given the limitations of the MRCI calculations. For example, size extensivity errors in truncated configuration interaction calculations can be somewhat compensated for by inclusion of the Davidson correction, but it is not always clear if this will improve the description of the excited state PESs~\citep{jt599,jt623}. Scalar relativistic effects can also improve the accuracy of the PES but to achieve a balanced description it is advisable to treat them together with other additional corrections to recover more of the correlation energy, e.g. the effects of core-correlation~\citep{jt599,jt623}, but these can be computationally expensive to include.

It is important to note that we intend to empirically refine the excited \A\ state PESs and this procedure will mitigate many of the errors associated with the underlying \textit{ab initio} calculations. Generally speaking, empirical refinement of a more accurate \textit{ab initio} PES will result in a more accurate final PES. However, the additional computational effort that is required to improve the \textit{ab initio} PES must be carefully considered, especially if there is sufficient experimental data available for refinement as is the case for CaOH. From our experience, purely \textit{ab initio} excited state PESs are never accurate enough for high-resolution spectroscopic applications and a degree of empirical refinement is always necessary.

\begin{figure}
\centering
\includegraphics{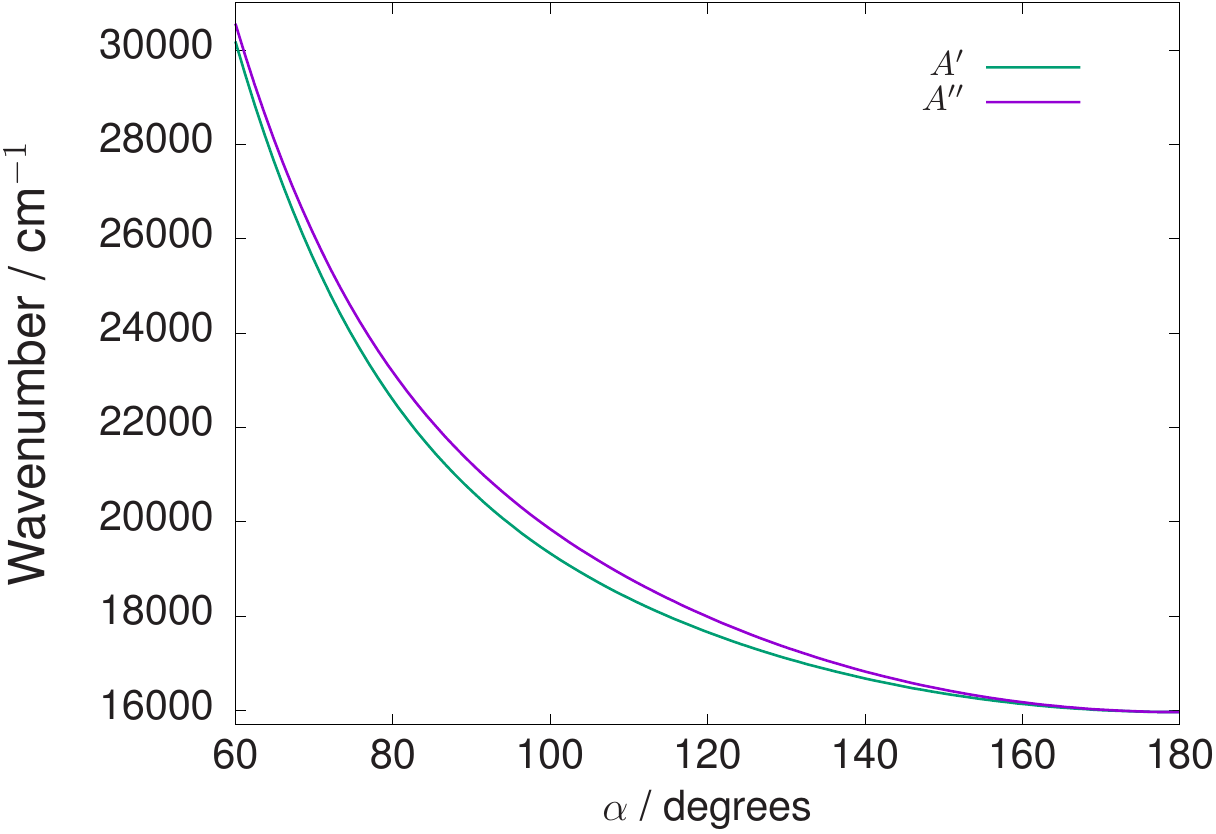}
\caption{\label{fig:A_state_bend}One-dimensional bending cut of the \A\ state potential energy surface. At bent configurations, when the bond-angle $\alpha<180^{\circ}$, the Renner-Teller effect lifts the degeneracy of the \A\ state, which splits into two separate components $A^{\prime}$ and $A^{\prime\prime}$.}
\end{figure}

As mentioned already, the Renner-Teller effect lifts the degeneracy of the \A\ state, which splits into two components $A^{\prime}$ and $A^{\prime\prime}$ at bent configurations, see Fig.~\ref{fig:A_state_bend}. The two surfaces were fitted independently using the analytic representation and coordinates given above, with the pure stretching parameters having identical values at the linear (equilibrium) configuration. For the $A^{\prime}$ component, 42 parameters (including 3 equilibrium parameters) were used in the fitting and reproduced the \textit{ab initio} data with a weighted rms error of 0.160~cm$^{-1}$. The fitting used Partridge and Schwenke's weighting function~\citep{97PaScxx.H2O} to favour energies closer to the excited state minimum and Watson's robust fitting scheme~\citep{03Watson.methods} to reduce the weights of outliers. When fitting the $A^{\prime\prime}$ component, only parameters associated with the bending coordinate were allowed to vary to ensure the correct Renner-Teller behaviour. Expansion parameters associated with stretching terms only were kept fixed to the values determined in the $A^{\prime}$ component fitting. For the $A^{\prime\prime}$ component, 42 parameters (including 3 equilibrium parameters) were used and the \textit{ab initio} data was reproduced with a weighted rms error of 0.196~cm$^{-1}$.

Spin-orbit coupling between the $A^{\prime}$ and $A^{\prime\prime}$ components of the \A\ state was also computed at the MRCI/aug-cc-pV5Z level of theory and fitted using the same representation and coordinates as the PESs. As the curve is much simpler in form, only a fourth-order expansion with 22 parameters (including 3 equilibrium parameters) was used, reproducing the \textit{ab initio} data  with a weighted rms error of 0.158~cm$^{-1}$.

A full empirical refinement of the \A\ state potentials will require development of the rovibronic code EVEREST~\citep{EVEREST}, which is beyond the scope of the present study. This will be carried out in future work before a comprehensive molecular line list of CaOH is computed for the ExoMol database~\citep{jt810,jt631,jt528}. To evaluate the quality of our calculations and PESs, we have adjusted the zeroth-order expansion parameters $f_{000}$ of the $A^{\prime}$ and $A^{\prime\prime}$ PESs and spin-orbit coupling curve to better agree with experiment. This can be understood as a shift in the minimum of the excited state PESs and a correction of the magnitude of the spin-orbit coupling to ensure accurate spin-orbit splitting in the energy level structure of CaOH. To achieve this, the two parameters (the $A^{\prime}$ and $A^{\prime\prime}$ PESs assume the same $f_{000}$ value) were adjusted in rovibronic calculations (discussed in Sec.~\ref{sec:everest}) to match the \A\ ground vibrational state $J=1.5$ energy levels at 15966.0948~cm$^{-1}$ ($e$ parity) and 16032.5536~cm$^{-1}$ ($f$ parity). Here, $e$ and $f$ are the Kronig parity quantum numbers~\citep{75BrHoHu.diatom}. These two energy levels were selected as they are the lowest energies available for the \A\ state in the MARVEL data set of empirical-quality CaOH term values~\citep{jt791}. No experimental values for the \A\ ground vibrational state $J=0.5$ energy levels are available.

\subsubsection{Dipole moment surfaces}

Transition DMSs for the $\tilde{A}$--$\tilde{X}$ band system were computed as expectation values at the MRCI/aug-cc-pV5Z level of theory on the same grid of nuclear geometries as the PES. To represent the transition DMSs analytically, we have adopted the $pq$ axis system~\citep{93JoJexx.H2O} and the following coordinates,
\begin{eqnarray}
\label{eq:coords_dms}
  \zeta_1 &=& r_1-r^{\rm eq}_1, \\
  \zeta_2 &=& r_2-r^{\rm eq}_2, \\
  \zeta_3 &=& \cos\alpha-\cos\alpha_{\rm eq},
\end{eqnarray}
where $r^{\rm eq}_1$, $r^{\rm eq}_2$, and $\alpha_{\rm eq}$ are the equilibrium values (see supplementary material for DMS parameters). For the perpendicular $\mu_x$ and $\mu_y$ components, the function
\begin{equation}
\label{eq:mu_p}
\mu_p =  \sum_{i_1,i_2,i_3} F^{(p)}_{i_1,i_2,i_3} \zeta_1^{i_1} \zeta_2^{i_2} \zeta_3^{i_3} ,
\end{equation}
 was used. While the parallel $\mu_z$ component was expressed as,
\begin{equation}
\label{eq:mu_q}
\mu_q =  \sin(\pi-\alpha)\sum_{i_1,i_2,i_3} F^{(q)}_{i_1,i_2,i_3} \zeta_1^{i_1} \zeta_2^{i_2} \zeta_3^{i_3} .
\end{equation}
A sixth-order expansion was used to represent the DMSs with the parameters $F^{(p/q)}_{i_1,i_2,i_3}$ determined in a least-squares fitting to the \textit{ab initio} data using Watson's robust fitting scheme~\citep{03Watson.methods}, and Partridge and Schwenke's weighting function~\citep{97PaScxx.H2O}. The $\mu_x$ and $\mu_y$ components each required 77 parameters (including the 3 equilibrium parameters) and reproduced the \textit{ab initio} data with a weighted rms error of $0.0012$ and $0.0013$~Debye, respectively. The $\mu_z$ component was fitted with 66 parameters (including 3 equilibrium parameters) and a weighted rms error of $0.0041$~Debye. All surfaces generated in this work are given as supplementary material along with programs to construct them.

\subsection{EVEREST calculations}
\label{sec:everest}

Rovibronic calculations were performed with the EVEREST code~\citep{EVEREST}, which is able to treat interacting states with spin-dependent coupling and Renner-Teller effects in triatomic molecules. EVEREST employs an exact kinetic energy operator and is general in its design, having been used in several spectroscopic studies on Renner-Teller systems such as the SiCCl radical~\citep{Mitrushchenkov2015}, the PCS radical~\citep{Finney2016} and the SiCN/SiNC system~\citep{Brites2013}.

Calculations for CaOH employed valence bond length-bond angle coordinates with a discrete variable representation (DVR) basis composed of 100 Sinc-DVR functions on both the Ca--O bond in the 2.6--7.0~$a_0$ interval and the O--H bond in the 1.1--6.0~$a_0$ interval, along with 120 Legendre functions for the $\angle({\rm CaOH})$ bond angle. Vibrational $J=0$ eigenfunctions with energies up to 10\,000~cm$^{-1}$ above the lowest vibronic state for $0\leq K\leq 27$, where $K=|\Lambda + l|$ ($\Lambda$ and $l$ are the projections of the electronic and vibrational angular momenta along the linear axis), were computed from a Hamiltonian with a dimension of 10\,000. For $K\geq 1$, the Renner-Teller effect was explicitly taken into account by solving the coupled $A^{\prime}/A^{\prime\prime}$ problem, see Ref.~\onlinecite{EVEREST} for details. The full rovibronic Hamiltonian including spin-orbit coupling was then built and diagonalized using these vibronic states for $J$ up to 125.5, where $J$ is the total angular momentum quantum number. The spin-orbit coupling curve, determined from \textit{ab initio} calculations described above, was represented as a fouth-order expansion using the same analytic representation and coordinates as the PESs. Computed rovibronic states were sufficiently converged with the chosen EVEREST calculation parameters, with convergence testing performed by increasing the dimension of the Hamiltonian, running calculations with $K\leq 50$, and using larger DVR grids for the stretching and bending modes.

\section{Results}
\label{sec:results}

There have been a number of experimental studies of the $\tilde{A}$--$\tilde{X}$ band system of CaOH~\citep{85BeBrxx.CaOH,91CoLiPr.CaOH,92LiCoxx.CaOH,92CoLiPr.CaOH,94CoLiPr.CaOH,95LiCoxx.CaOH,06DiShWa.CaOH} but only four vibrational states of the \A\ electronic state have been characterised; $(0,0,0)$, $(1,0,0)$, $(0,1^1,0)$ and $(0,2^0,0)$. The reader is referred to Figure 1 of Ref.~\onlinecite{jt791} for an energy level diagram of the experimentally measured states of CaOH. Here, vibrational states are assigned using normal mode notation $(v_1,v_2^{L},v_3)$, where $v_1$ and $v_3$ correspond to the symmetric and asymmetric stretching modes, respectively, and $v_2$ labels the bending mode. The quantum number $L$ is related to the absolute value of the vibrational angular momentum quantum number $l$ associated with the $\nu_2$ bending mode, $L=|l|$, with the vibrational quantum number $l$ taking the values $|l| = v_2, v_2-2, v_2-4, \ldots, 0\,({\rm or}\,1)$. Several quantum numbers are required to uniquely label the rovibronic states of CaOH including the total angular momentum quantum number $J$, the rotationless parity $e/f$, and the quantum labels $F_1$ and $F_2$ denoting spin components $J = N + 1/2$ and $J = N - 1/2$, respectively, where $N$ is the rotational angular momentum quantum number. Couplings also give rise to additional labelling, for example, in the $(0,1^1,0)$ vibrational state, coupling between the vibrational angular momentum with $l=\pm 1$ and the electronic angular momentum with $\Lambda = \pm 1$ leads to the three vibronic components $|\Lambda+l| = 0,0,2$, i.e. $\Sigma^{+}$, $\Sigma^{-}$ and a doubly degenerate state $\Delta$.  These vibronic states are then assigned $\tilde{A} (0,1^1,0)$ $^2\Sigma^{\pm}$ and  $\tilde{A}(0,1^1,0)$ $^2\Delta$. Furthermore, in Renner-Teller split states involving the $\nu_2$ bending mode, $\mu$ and $\kappa$ label the lower and upper levels, respectively, of the pair of states associated with $K>0$.

\begin{figure}
\centering
\includegraphics{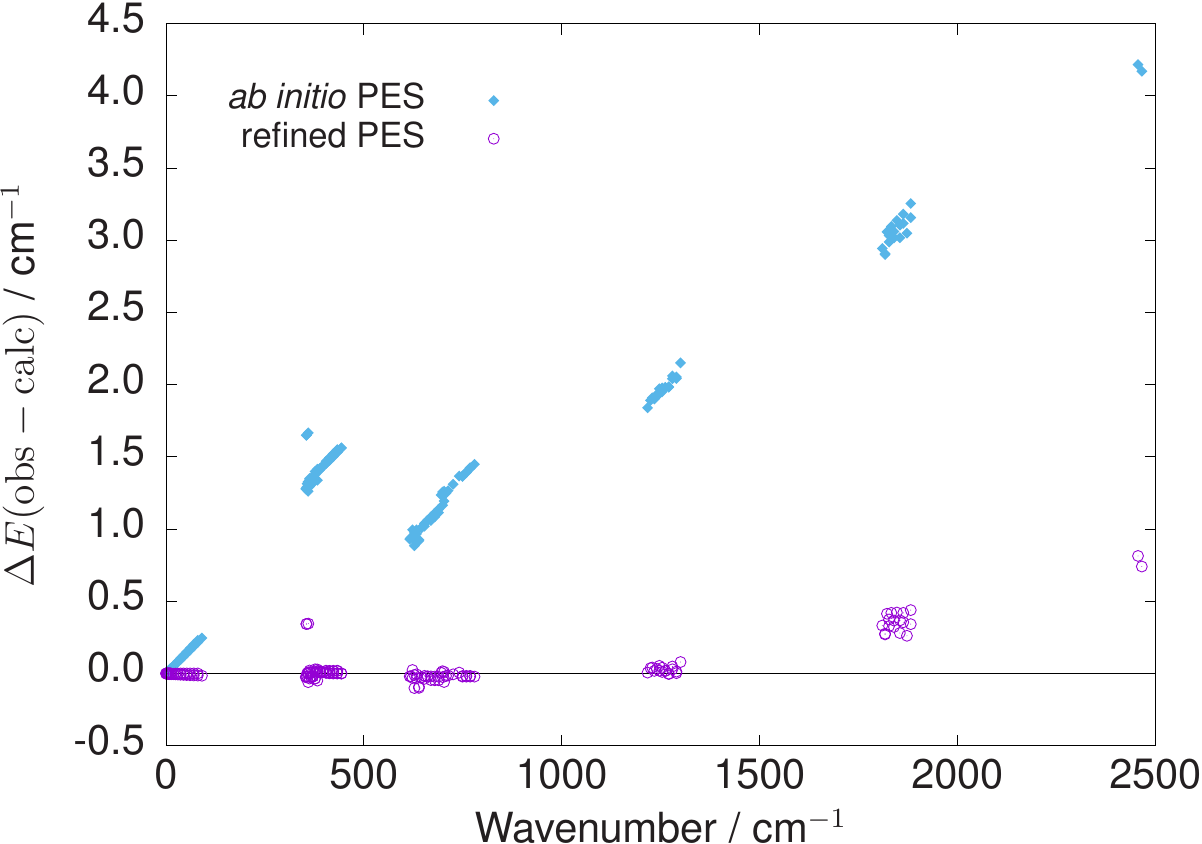}
\caption{\label{fig:res_x2sigma}Residual errors $\Delta E (\rm obs-calc)$ between the observed and calculated \X\ term values of CaOH up to $J=15.5$ using the \textit{ab initio} and empirically refined PESs (see Table~\ref{tab:caoh_en_x2sigma} for values).}
\end{figure}

The results of the \X\ state PES refinement are shown in Fig.~\ref{fig:res_x2sigma} and Table~\ref{tab:caoh_en_x2sigma}. Only results up to $J=4.5$ are listed in Table~\ref{tab:caoh_en_x2sigma} as the residual errors $\Delta E({\rm obs-calc})$ between the observed and calculated values up to $J=15.5$ assume consistent values for the different vibrational states, as seen in Fig.~\ref{fig:res_x2sigma}. The original \textit{ab initio} PES reproduces the 162 term values up to $J=15.5$ with an rms error of 1.17~cm$^{-1}$, while for the refined PES the rms error is substantially reduced to 0.06~cm$^{-1}$. Closer inspection of Fig.~\ref{fig:res_x2sigma} and Table~\ref{tab:caoh_en_x2sigma} reveals several small outliers associated with low-$J$ values of the $(0,1^1,0)$ state around $\approx 355$~cm$^{-1}$. We believe this to be an issue with the experimental values since the higher-$J$ values in the same vibrational state are reproduced much better, with residual errors close to zero. Low-$J$ states are usually more difficult to characterise in experiment as there are fewer transitions linking them, and this was certainly the case for the states in question as determined in the MARVEL analysis of CaOH~\citep{jt791}. The original measurements of the low-$J$ values of the $(0,1^1,0)$ state are from Refs.~\onlinecite{94CoLiPr.CaOH,95LiCoxx.CaOH}, while further values for this state with $12.5 \leq J \leq 19.5$ were also measured in Ref.~\onlinecite{96ZiFlAn.CaOH}.

\begin{table}
\caption{\label{tab:caoh_en_x2sigma}Calculated \X\ energy levels of CaOH using the \textit{ab initio} and empirically-refined PESs compared against empirically-derived values~\citep{jt791} (also illustrated in Fig.~\ref{fig:res_x2sigma} up to $J=15.5$).}
\setlength{\tabcolsep}{8pt}
\begin{tabular}{cccccccrrrrr}
\hline\hline
$J$	&	$e/f$	&	$v_1$	&	$v_2$	&	$L$	&	$v_3$	&	$F_1/F_2$	&	Observed	&	\textit{ab initio}	&	obs$-$calc (ai) & refined & obs$-$calc (ref)	\\
\hline
0.5	&	$e$	&	0	&	0	&	0	&	0	&	$F_1$	&	0.00	&	0.00	&	0.00	&	0.00	&	0.00	\\[-3mm]
0.5	&	$f$	&	0	&	0	&	0	&	0	&	$F_2$	&	0.67	&	0.67	&	0.00	&	0.67	&	0.00	\\[-3mm]
1.5	&	$e$	&	0	&	0	&	0	&	0	&	$F_1$	&	0.67	&	0.67	&	0.00	&	0.67	&	0.00	\\[-3mm]
1.5	&	$f$	&	0	&	0	&	0	&	0	&	$F_2$	&	2.00	&	2.00	&	0.00	&	2.01	&	0.00	\\[-3mm]
1.5	&	$e$	&	0	&	1	&	1	&	0	&	$F_1$	&	353.22	&	351.95	&	1.28	&	353.30	&	-0.08	\\[-3mm]
1.5	&	$e$	&	0	&	1	&	1	&	0	&	$F_2$	&	354.56	&	353.28	&	1.28	&	354.64	&	-0.08	\\[-3mm]
1.5	&	$f$	&	0	&	1	&	1	&	0	&	$F_2$	&	354.92	&	353.27	&	1.65	&	354.63	&	0.29	\\[-3mm]
2.5	&	$e$	&	0	&	0	&	0	&	0	&	$F_1$	&	2.01	&	2.00	&	0.01	&	2.01	&	0.00	\\[-3mm]
2.5	&	$f$	&	0	&	0	&	0	&	0	&	$F_2$	&	4.01	&	4.00	&	0.01	&	4.01	&	0.00	\\[-3mm]
2.5	&	$f$	&	0	&	1	&	1	&	0	&	$F_1$	&	354.56	&	353.28	&	1.28	&	354.64	&	-0.08	\\[-3mm]
2.5	&	$e$	&	0	&	1	&	1	&	0	&	$F_1$	&	354.92	&	353.27	&	1.65	&	354.63	&	0.29	\\[-3mm]
3.5	&	$e$	&	0	&	0	&	0	&	0	&	$F_1$	&	4.01	&	4.00	&	0.01	&	4.01	&	0.00	\\[-3mm]
3.5	&	$f$	&	0	&	0	&	0	&	0	&	$F_2$	&	6.68	&	6.67	&	0.02	&	6.69	&	0.00	\\[-3mm]
3.5	&	$e$	&	0	&	1	&	1	&	0	&	$F_1$	&	356.56	&	355.27	&	1.29	&	356.63	&	-0.07	\\[-3mm]
3.5	&	$f$	&	0	&	1	&	1	&	0	&	$F_1$	&	356.59	&	355.28	&	1.31	&	356.64	&	-0.05	\\[-3mm]
3.5	&	$e$	&	0	&	1	&	1	&	0	&	$F_2$	&	359.20	&	357.94	&	1.26	&	359.31	&	-0.11	\\[-3mm]
3.5	&	$f$	&	0	&	1	&	1	&	0	&	$F_2$	&	359.25	&	357.92	&	1.33	&	359.30	&	-0.04	\\[-3mm]
4.5	&	$e$	&	0	&	0	&	0	&	0	&	$F_1$	&	6.69	&	6.67	&	0.02	&	6.69	&	0.00	\\[-3mm]
4.5	&	$f$	&	0	&	0	&	0	&	0	&	$F_2$	&	10.03	&	10.00	&	0.02	&	10.03	&	0.00	\\[-3mm]
4.5	&	$e$	&	0	&	1	&	1	&	0	&	$F_1$	&	359.24	&	357.92	&	1.32	&	359.30	&	-0.05	\\[-3mm]
4.5	&	$f$	&	0	&	1	&	1	&	0	&	$F_1$	&	359.60	&	357.94	&	1.67	&	359.31	&	0.29	\\[-3mm]
4.5	&	$f$	&	0	&	1	&	1	&	0	&	$F_2$	&	362.58	&	361.25	&	1.34	&	362.63	&	-0.04	\\[-3mm]
4.5	&	$e$	&	1	&	0	&	0	&	0	&	$F_1$	&	615.67	&	614.74	&	0.93	&	615.68	&	-0.02	\\[-3mm]
4.5	&	$e$	&	0	&	2	&	0	&	0	&	$F_1$	&	695.36	&	694.12	&	1.24	&	695.34	&	0.01	\\[-3mm]
4.5	&	$e$	&	2	&	0	&	0	&	0	&	$F_1$	&	1216.72	&	1214.88	&	1.84	&	1216.81	&	-0.09	\\[-3mm]
4.5	&	$e$	&	3	&	0	&	0	&	0	&	$F_1$	&	1809.60	&	1806.65	&	2.94	&	1809.62	&	-0.02	\\
\hline\hline
\end{tabular}
\end{table}

\begin{figure}
\centering
\includegraphics{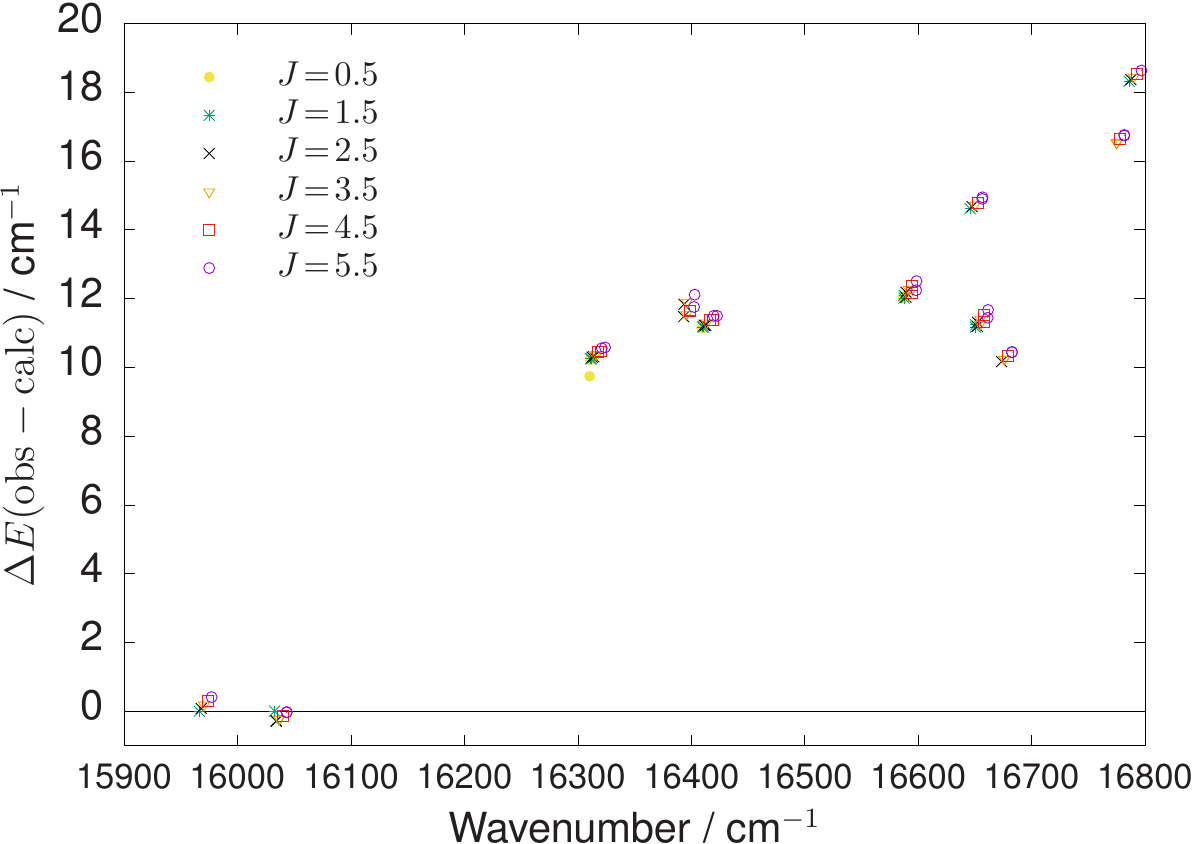}
\caption{\label{fig:res_caoh}Residual errors $\Delta E (\rm obs-calc)$ between the observed and calculated \A\ term values of CaOH up to $J=5.5$ (see Table~\ref{tab:caoh_en} for values).}
\end{figure}

Table~\ref{tab:caoh_en} and Fig.~\ref{fig:res_caoh} compare the computed rovibronic energy levels of the \A\ state  against all available empirically-derived energy levels~\citep{jt791}. Again, only results up to $J=2.5$ are listed in Table~\ref{tab:caoh_en} as the residual errors $\Delta E({\rm obs-calc})$ between the observed and calculated values up to $J=5.5$ assume consistent values for the different vibrational states, as seen in Fig.~\ref{fig:res_caoh}. As discussed earlier, because the zeroth-order parameters of the $A^{\prime}$ and $A^{\prime\prime}$ PESs and the spin-orbit coupling curve were adjusted to match the \A\ ground vibrational state $J=1.5$ energy levels at 15966.0948~cm$^{-1}$ ($e$ parity) and 16032.5536~cm$^{-1}$ ($f$ parity), the residual errors for all the ground vibrational state term values are close to zero. All other computed rovibronic energy levels are underestimated with residual errors ranging from 10--20~cm$^{-1}$. This is to be expected given that the excited state PESs are essentially \textit{ab initio} surfaces that have not been subject to rigorous empirical refinement. Doing so would considerably reduce the errors by at least an order-of-magnitude or better, and this work will be undertaken in the future before a new high-temperature line list is generated for CaOH.

\begin{table}
\caption{\label{tab:caoh_en}Calculated \A\ energy levels of CaOH compared against empirically-derived values~\citep{jt791} (also illustrated in Fig.~\ref{fig:res_caoh} up to $J=5.5$).}
\setlength{\tabcolsep}{8pt}
\begin{tabular}{lcccccccccr}
\hline\hline
Vibronic state	&	$J$	&	$e/f$	&	$v_1$	&	$v_2$	&	$L$	&	$v_3$	&	$F_1/F_2$	&	Observed	&	Calculated	&	obs$-$calc	\\
\hline
$\mu\tilde{A}\,^2\Sigma$	&	0.5	&	$e$	&	0	&	1	&	1	&	0	&	$F_1$	&	16310.23	&	16300.01	&	10.22	\\[-3mm]
$\mu\tilde{A}\,^2\Sigma$	&	0.5	&	$f$	&	0	&	1	&	1	&	0	&	$F_2$	&	16310.71	&	16300.48	&	10.22	\\[-3mm]
$\kappa\tilde{A}\,^2\Sigma$	&	0.5	&	$f$	&	0	&	1	&	1	&	0	&	$F_1$	&	16408.92	&	16397.77	&	11.15	\\[-3mm]
$\kappa\tilde{A}\,^2\Sigma$	&	0.5	&	$e$	&	0	&	1	&	1	&	0	&	$F_2$	&	16409.39	&	16398.25	&	11.14	\\[-3mm]
$\tilde{A}\,^2\Pi$	&	0.5	&	$f$	&	1	&	0	&	0	&	0	&	$F_1$	&	16586.39	&	16574.38	&	12.01	\\[-3mm]
$\tilde{A}\,^2\Pi$	&	0.5	&	$e$	&	1	&	0	&	0	&	0	&	$F_1$	&	16586.44	&	16574.38	&	12.05	\\[-3mm]
$\tilde{A}\,^2\Pi$	&	1.5	&	$e$	&	0	&	0	&	0	&	0	&	$F_1$	&	15966.10	&	15966.10	&	0.00	\\[-3mm]
$\tilde{A}\,^2\Pi$	&	1.5	&	$f$	&	0	&	0	&	0	&	0	&	$F_2$	&	16032.56	&	16032.56	&	0.00	\\[-3mm]
$\mu\tilde{A}\,^2\Sigma$	&	1.5	&	$e$	&	0	&	1	&	1	&	0	&	$F_1$	&	16311.01	&	16300.75	&	10.26	\\[-3mm]
$\mu\tilde{A}\,^2\Sigma$	&	1.5	&	$f$	&	0	&	1	&	1	&	0	&	$F_2$	&	16311.96	&	16301.71	&	10.25	\\[-3mm]
$\kappa\tilde{A}\,^2\Sigma$	&	1.5	&	$f$	&	0	&	1	&	1	&	0	&	$F_1$	&	16409.70	&	16398.52	&	11.18	\\[-3mm]
$\kappa\tilde{A}\,^2\Sigma$	&	1.5	&	$e$	&	0	&	1	&	1	&	0	&	$F_2$	&	16410.66	&	16399.48	&	11.18	\\[-3mm]
$\tilde{A}\,^2\Pi$	&	1.5	&	$f$	&	1	&	0	&	0	&	0	&	$F_1$	&	16587.38	&	16575.37	&	12.02	\\[-3mm]
$\tilde{A}\,^2\Pi$	&	1.5	&	$e$	&	1	&	0	&	0	&	0	&	$F_1$	&	16587.46	&	16575.37	&	12.10	\\[-3mm]
$\tilde{A}\,^2\Pi$	&	1.5	&	$e$	&	1	&	0	&	0	&	0	&	$F_2$	&	16645.70	&	16631.08	&	14.62	\\[-3mm]
$\mu\tilde{A}\,^2\Pi$	&	1.5	&	$f$	&	0	&	2	&	0	&	0	&	$F_1$	&	16650.41	&	16639.24	&	11.17	\\[-3mm]
$\mu\tilde{A}\,^2\Pi$	&	1.5	&	$e$	&	0	&	2	&	0	&	0	&	$F_1$	&	16650.49	&	16639.25	&	11.24	\\[-3mm]
$\kappa\tilde{A}\,^2\Pi$	&	1.5	&	$e$	&	0	&	2	&	0	&	0	&	$F_1$	&	16785.75	&	16767.42	&	18.32	\\[-3mm]
$\tilde{A}\,^2\Pi$	&	2.5	&	$e$	&	0	&	0	&	0	&	0	&	$F_1$	&	15967.82	&	15967.75	&	0.08	\\[-3mm]
$\tilde{A}\,^2\Pi$	&	2.5	&	$f$	&	0	&	0	&	0	&	0	&	$F_2$	&	16033.94	&	16034.22	&	-0.29	\\[-3mm]
$\tilde{A}\,^2\Pi$	&	2.5	&	$e$	&	0	&	0	&	0	&	0	&	$F_2$	&	16033.94	&	16034.22	&	-0.28	\\[-3mm]
$\mu\tilde{A}\,^2\Sigma$	&	2.5	&	$e$	&	0	&	1	&	1	&	0	&	$F_1$	&	16312.46	&	16302.16	&	10.30	\\[-3mm]
$\mu\tilde{A}\,^2\Sigma$	&	2.5	&	$f$	&	0	&	1	&	1	&	0	&	$F_2$	&	16313.91	&	16303.60	&	10.31	\\[-3mm]
$\tilde{A}\,^2\Delta$	&	2.5	&	$e$	&	0	&	1	&	1	&	0	&	$F_2$	&	16393.14	&	16381.66	&	11.48	\\[-3mm]
$\tilde{A}\,^2\Delta$	&	2.5	&	$f$	&	0	&	1	&	1	&	0	&	$F_2$	&	16393.50	&	16381.66	&	11.84	\\[-3mm]
$\kappa\tilde{A}\,^2\Sigma$	&	2.5	&	$f$	&	0	&	1	&	1	&	0	&	$F_1$	&	16411.16	&	16399.93	&	11.23	\\[-3mm]
$\kappa\tilde{A}\,^2\Sigma$	&	2.5	&	$e$	&	0	&	1	&	1	&	0	&	$F_2$	&	16412.59	&	16401.37	&	11.22	\\[-3mm]
$\tilde{A}\,^2\Pi$	&	2.5	&	$f$	&	1	&	0	&	0	&	0	&	$F_1$	&	16589.04	&	16577.00	&	12.04	\\[-3mm]
$\tilde{A}\,^2\Pi$	&	2.5	&	$e$	&	1	&	0	&	0	&	0	&	$F_1$	&	16589.19	&	16577.00	&	12.19	\\[-3mm]
$\tilde{A}\,^2\Pi$	&	2.5	&	$e$	&	1	&	0	&	0	&	0	&	$F_2$	&	16647.40	&	16632.72	&	14.68	\\[-3mm]
$\mu\tilde{A}\,^2\Pi$	&	2.5	&	$f$	&	0	&	2	&	0	&	0	&	$F_1$	&	16652.08	&	16640.89	&	11.19	\\[-3mm]
$\mu\tilde{A}\,^2\Pi$	&	2.5	&	$e$	&	0	&	2	&	0	&	0	&	$F_1$	&	16652.21	&	16640.89	&	11.31	\\[-3mm]
$\mu\tilde{A}\,^2\Pi$	&	2.5	&	$f$	&	0	&	2	&	0	&	0	&	$F_2$	&	16673.42	&	16663.25	&	10.17	\\[-3mm]
$\kappa\tilde{A}\,^2\Pi$	&	2.5	&	$e$	&	0	&	2	&	0	&	0	&	$F_1$	&	16787.47	&	16769.10	&	18.37	\\
\hline\hline
\end{tabular}
\end{table}

\begin{figure}
\centering
\includegraphics{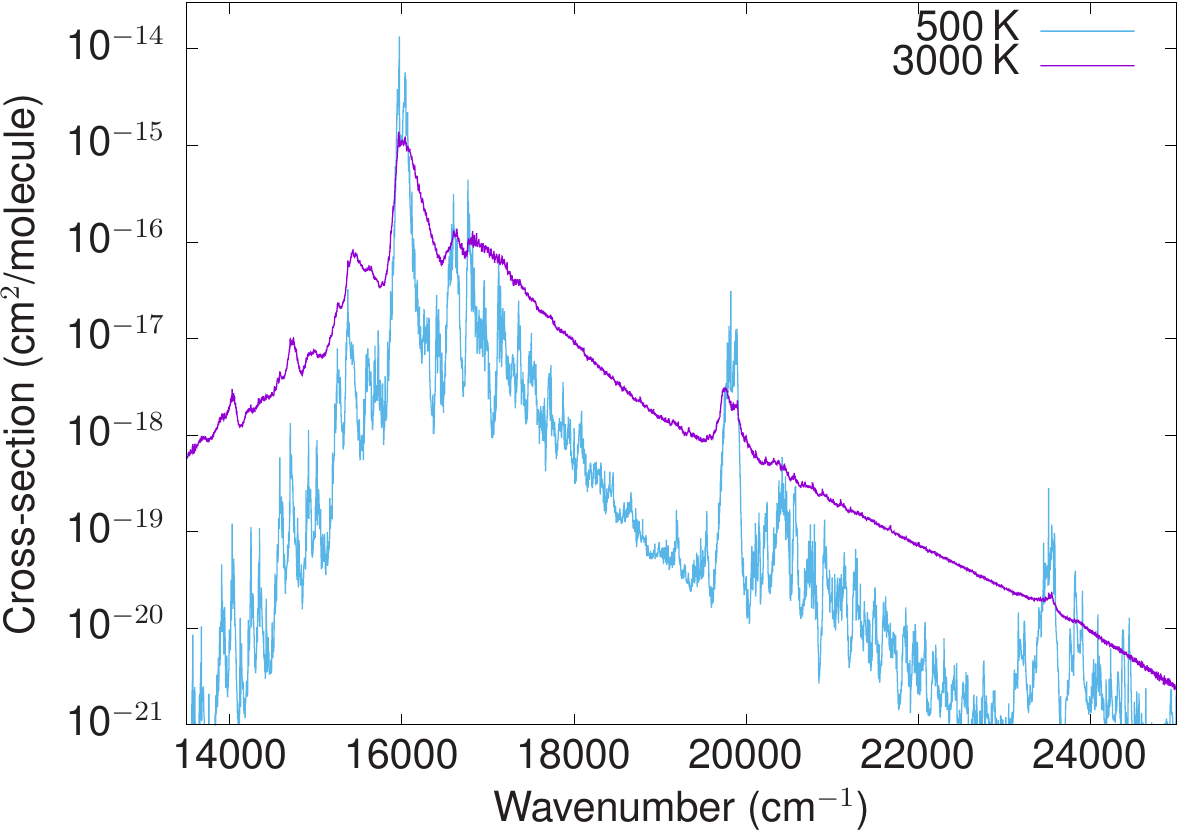}
\caption{\label{fig:caoh_spectrum}Absorption cross-sections of CaOH at $T=500$~K and 3000~K, simulated at a resolution of 1~cm$^{-1}$ and modelled with a Gaussian line profile with a half width at half maximum (HWHM) of 1~cm$^{-1}$.}
\end{figure}

\begin{figure}
\centering
\includegraphics[width=0.49\textwidth]{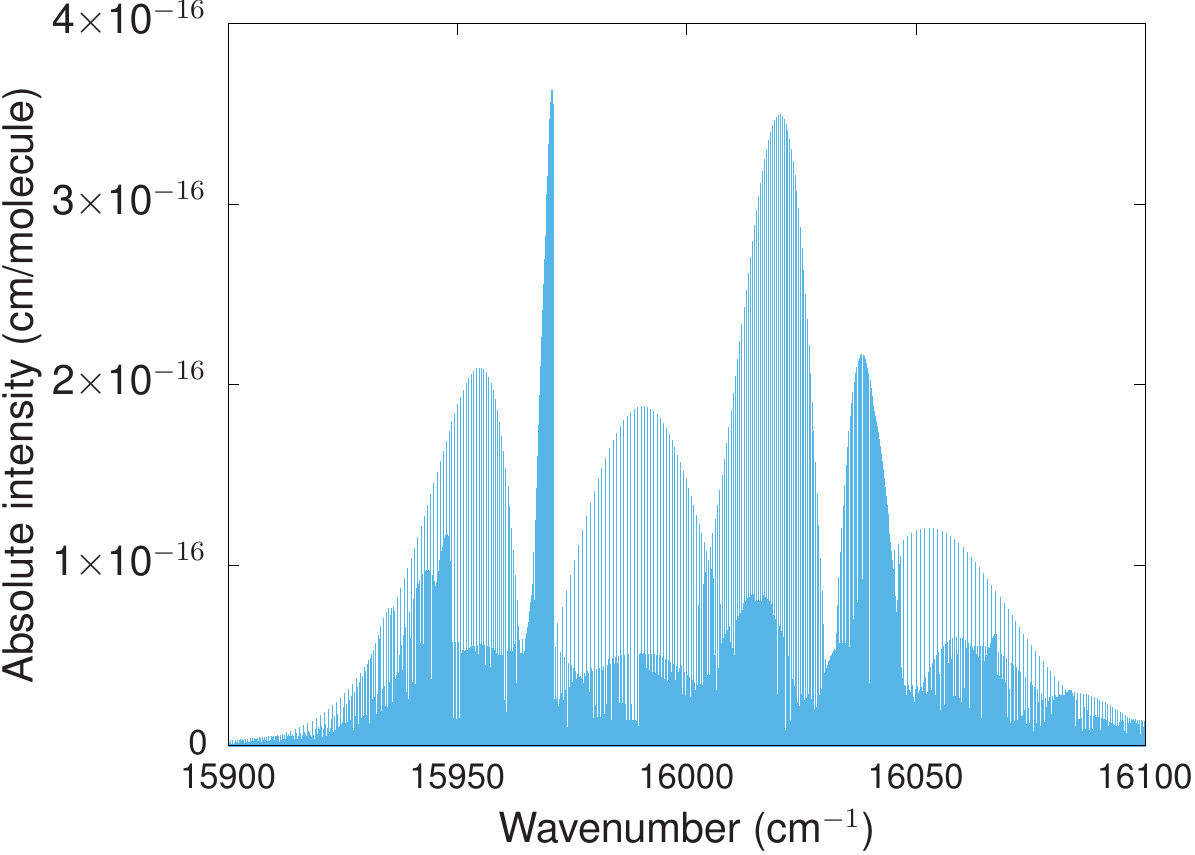}
\includegraphics[width=0.49\textwidth]{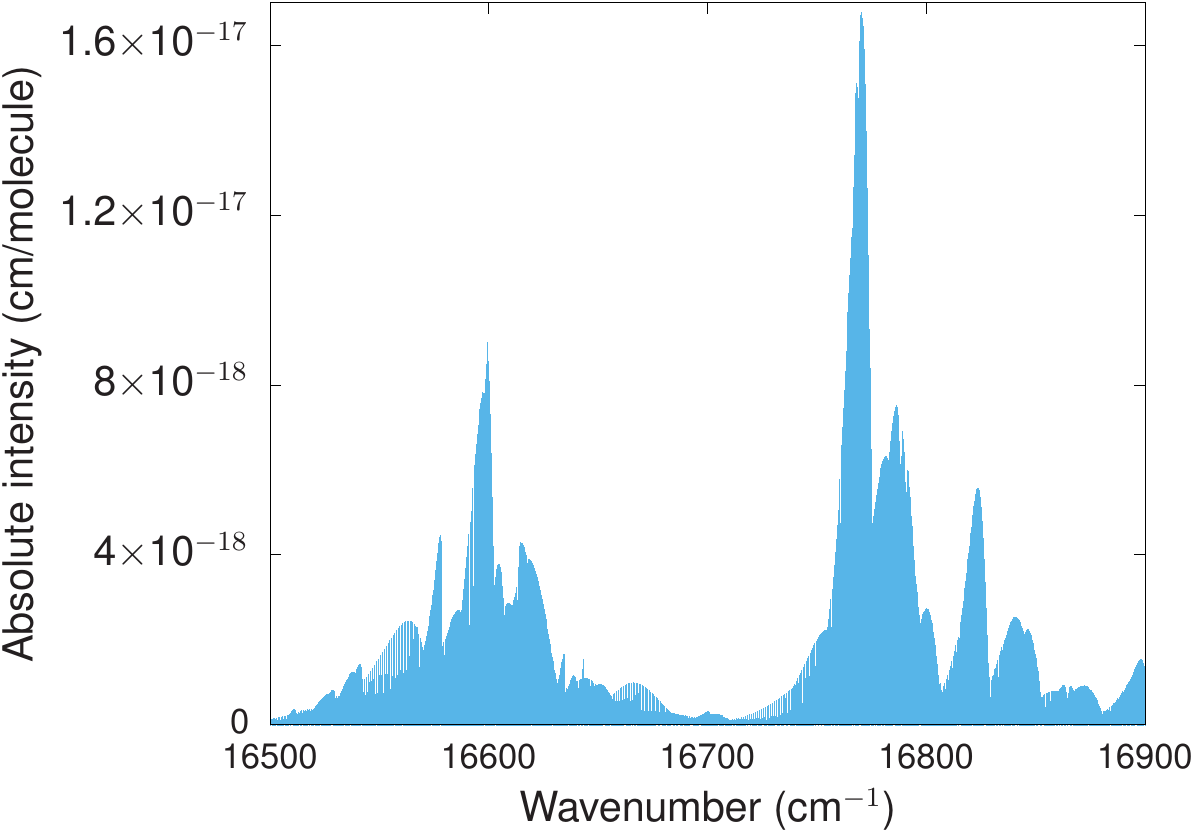}
\caption{\label{fig:caoh_stick}Absolute line intensities of the strongest \A\ bands of CaOH at a temperature of $T=700$~K.}
\end{figure}

An overview of the spectrum of the $\tilde{A}$--$\tilde{X}$ band system is shown in Fig.~\ref{fig:caoh_spectrum}, where we have simulated absolute absorption cross-sections at a resolution of 1~cm$^{-1}$ using a Gaussian line profile with a half width at half maximum (HWHM) of 1~cm$^{-1}$. The strongest rovibronic features occur around $\approx 15\,950$~cm$^{-1}$ and this band still dominates the CaOH spectrum at high temperatures ($T=3000$~K), which is significant for any future observation of this molecule in exoplanetary atmospheres, particularly hot rocky planets. Absolute absorption line intensities at $T=700$~K have been plotted in Fig.~\ref{fig:caoh_stick} to illustrate the structure of this region. Since there is very little intensity information available on the $\tilde{A}$--$\tilde{X}$ band system we have been unable to properly compare this with any measured spectra. However, the computed spectra in Fig.~\ref{fig:caoh_stick} do not exhibit the typical band head structure associated with electronic spectra and we suspect this is due to the quality of the underlying \textit{ab initio} PES used in our calculations. Studies in the ExoMol group have shown that rovibronic spectra can be very sensitive to the quality of the underlying PESs, and that the ``cliff-like'' electronic band heads usually emerge after empirical refinement of the excited state PESs~\citep{jt774}.

\begin{figure}
\centering
\includegraphics[width=0.49\textwidth]{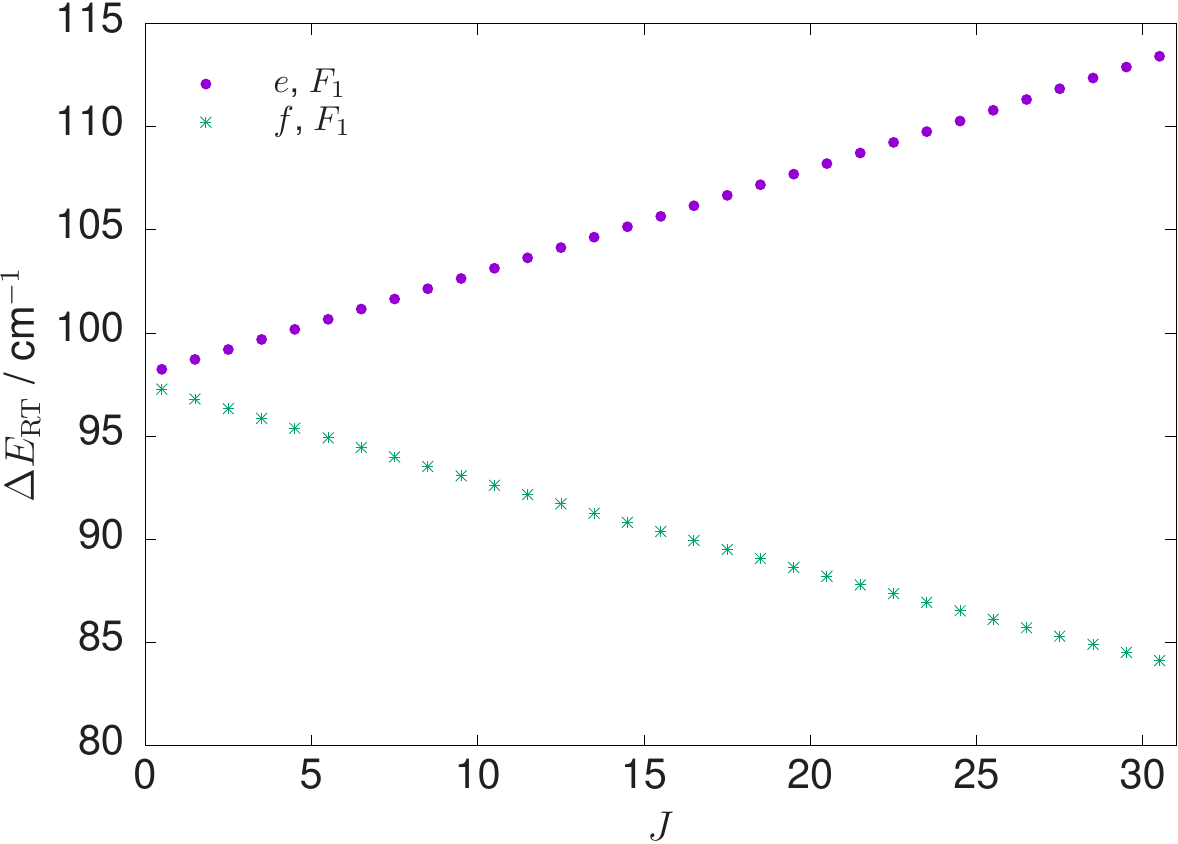}
\includegraphics[width=0.49\textwidth]{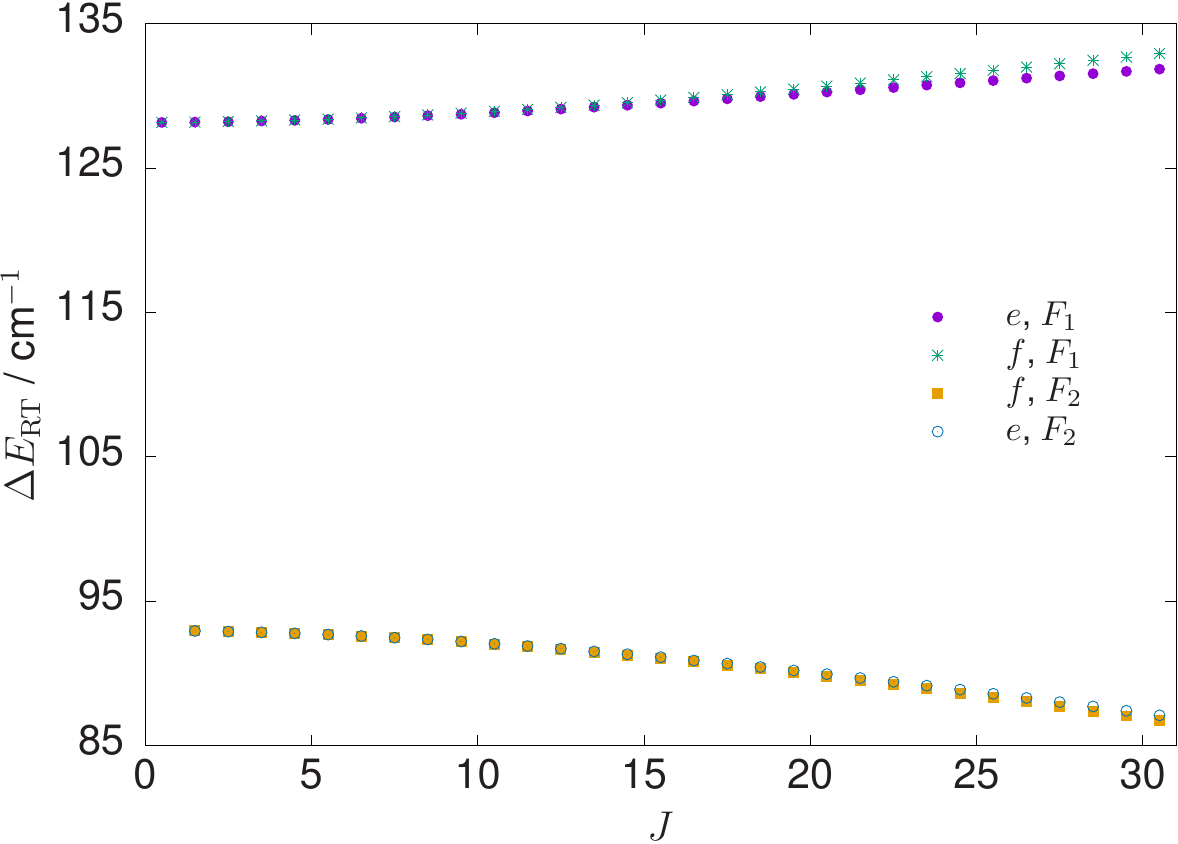}
\caption{\label{fig:rt_splitting}Size of the Renner-Teller splitting $\Delta E_{\rm RT}$ with total angular momentum quantum number $J$ for the $(0,1^1,0)$ (left panel) and $(0,2^0,0)$ (right panel) vibrational states. Here, the labels $e/f$ refer to the rotationless parity and $F_1/F_2$ the different spin components (see text).}
\end{figure}

In Fig.~\ref{fig:rt_splitting}, the size of the Renner-Teller splitting $\Delta E_{\rm RT}$ for the $(0,1^1,0)$ and $(0,2^0,0)$ vibrational states is plotted up to $J=30.5$. Interestingly, in the $(0,1^1,0)$ state where the absolute value of the vibrational angular momentum quantum number $|l|=1$, the size of the Renner-Teller splitting can be seen to change with $J$, increasing between the $e$ parity levels and decreasing for the $f$ parity levels. For the $(0,2^0,0)$ state where $|l|=0$, the size of the Renner-Teller splitting remains more or less constant with $J$.

\section{Conclusions}
\label{sec:conc}

The \A--\X\ rovibronic spectrum of CaOH has been studied using a combination of \textit{ab initio} theory and variational nuclear motion calculations. This band system is of great interest to the exoplanet community as the CaOH radical is expected to be present in the atmospheres of hot rocky super-Earth exoplanets but a lack of spectroscopic data is hampering its detection. New PESs for the \X\ and \A\ states have been produced along with $\tilde{A}$--$\tilde{X}$ transition DMSs and \A\ spin-orbit coupling. The ground \X\ state PES, originally a high-level \textit{ab initio} surface from Koput and Peterson~\citep{02KoPexx.CaOH}, was empirically refined to significantly improve its accuracy, reproducing all available observed rovibrational term values up to $J=15.5$ with an rms error of 0.06~cm$^{-1}$. For the excited \A\ state PESs, only the zeroth-order parameters were empirically tuned alongside the spin-orbit coupling curve. Computed \A\ term values for the excited vibrational states displayed errors of 10--20~cm$^{-1}$ compared to empirically-derived energies. Variational calculations utilised the nuclear motion code EVEREST~\citep{EVEREST}, which is able to treat the Renner-Teller effect and accurately predict Renner-Teller splittings in the excited bending states of CaOH. Preliminary line list calculations included transitions up to $J=125.5$ and the resulting spectra showed clear band structure with the strongest features occurring around the $\approx 15\,950$~cm$^{-1}$ region, which still dominates the CaOH spectrum at high temperatures. In future work we intend to perform a full empirical refinement of the \A\ state PESs, which will considerably improve the accuracy of our spectroscopic model. A comprehensive molecular line list of CaOH will then be generated for inclusion in the ExoMol database~\citep{jt810,jt631,jt528}.

\section*{Supplementary Material}

See the supplementary material for the expansion parameters and Fortran routines to construct the PESs, DMSs and spin-orbit coupling curve of CaOH.

\begin{acknowledgments}
This work was supported by the STFC Projects No. ST/M001334/1 and ST/R000476/1. The authors acknowledge the use of the UCL Legion High Performance Computing Facility (Legion@UCL) and associated support services in the completion of this work, along with the Cambridge Service for Data Driven Discovery (CSD3), part of which is operated by the University of Cambridge Research Computing on behalf of the STFC DiRAC HPC Facility (www.dirac.ac.uk). The DiRAC component of CSD3 was funded by BEIS capital funding via STFC capital grants ST/P002307/1 and ST/R002452/1 and STFC operations grant ST/R00689X/1. DiRAC is part of the National e-Infrastructure.
This work was also supported by the European
Research Council (ERC) under the European Union’s Horizon 2020 research and innovation
programme through Advance Grant number 883830.
\end{acknowledgments}

\section*{Data availability}
The data that support the findings of this study are available within the article and its supplementary material.

%

\end{document}